\documentclass[aps,prl,twocolumn,superscriptaddress,amsmath,amssymb]{revtex4}
\usepackage{float}
\usepackage{makeidx}
\usepackage{graphicx,amsmath}
\usepackage{CJK}
\usepackage{colordvi}
\usepackage{color}

\begin{document}
\begin{CJK*}{GB}{gbsn}

\title{Spin-Orbit Coupling Induced Coherent Production of Feshbach Molecules in a Degenerate Fermi Gas}
\author{Zhengkun Fu}
\affiliation{State Key Laboratory of Quantum Optics and Quantum
Optics Devices, Institute of Opto-Electronics, Shanxi University,
Taiyuan 030006, P.R.China }
\author{Lianghui Huang}
\affiliation{State Key Laboratory of Quantum Optics and Quantum
Optics Devices, Institute of Opto-Electronics, Shanxi University,
Taiyuan 030006, P.R.China }
\author{Zengming Meng}
\affiliation{State Key Laboratory of Quantum Optics and Quantum
Optics Devices, Institute of Opto-Electronics, Shanxi University,
Taiyuan 030006, P.R.China }
\author{Pengjun Wang}
\affiliation{State Key Laboratory of Quantum Optics and Quantum
Optics Devices, Institute of Opto-Electronics, Shanxi University,
Taiyuan 030006, P.R.China }
\author{Long Zhang}
\affiliation{Hefei National Laboratory for Physical Sciences at
Microscale and Department of Modern Physics, University of Science
and Technology of China, Hefei, Anhui 230026, P.R.China}
\author{Shizhong Zhang}
\affiliation{Department of Physics and Center of Theoretical and Computational Physics, The University of Hong Kong, Hong Kong, China}
\author{Hui Zhai}
\affiliation{Instiute for Advanced Study, Tsinghua University,
Beijing, 100084, P.R.China}
\author{Peng Zhang}
\affiliation{Department of Physics, Renmin University of China,
Beijing, 100872, P.R.China}
\author{Jing Zhang}
\email{jzhang74@yahoo.com; jzhang74@sxu.edu.cn}
\affiliation{State Key Laboratory of Quantum Optics and Quantum Optics Devices,
Institute of Opto-Electronics, Shanxi University, Taiyuan 030006,
P.R.China }

\begin{abstract}

\end{abstract}

\maketitle
\end{CJK*}

\textbf{Searching for topological superconductors is a challenging
task in physics nowadays. One of the most promising scheme is by
utilizing spin-orbit (SO) coupling, a spin polarized metal proximate
to an $s$-wave superconductor can exhibit $p$-wave
pairing~\cite{kane}. Recently, synthetic SO couplings have also been
realized by two-photon Raman process in
bosonic~\cite{spielman,Jing_PRA,spielman2,Shuai1,Shuai2,Shuai3,Washington,spielman_ZB}
and fermonic~\cite{Jing, MIT,Jing_molecule,Spielman-Fermi} cold atom
systems. In cold atom systems, instead of proximity effect, $s$-wave
pairing force originates from Feshbach resonance, in which an
$s$-wave Feshbach molecular state is tuned to scattering threshold
and resonantly couples to itinerant atoms~\cite{FR}. In this work we
demonstrate a dynamic process in which SO coupling can coherently
produce $s$-wave Feshbach molecules from a fully polarized Fermi
gas, and can induce a coherent oscillation between Feshbach
molecules and spin polarized gas. For comparison, we also show that
such phenomena are absent if the inter-component coupling is
momentum-independent. This demonstrates experimentally that SO
coupling does provide finite matrix element between a singlet state
and a triplet state, and therefore, implies the bound pairs of a
system with SO coupling have triplet $p$-wave component, which can
become topological superfluid by further cooling these pairs to
condensation and confining them to lower dimension.}

\begin{figure}[tbp]
\includegraphics[width=0.3\textwidth]{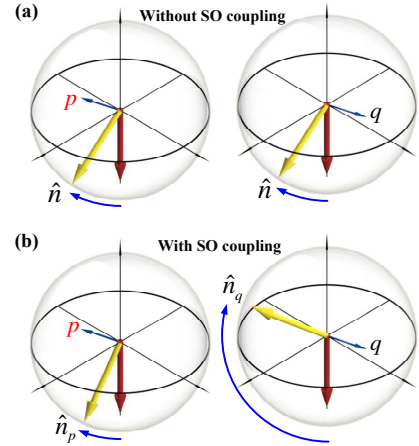}
\caption{(Color online). \textbf{Schematic of how SO coupling can
induce transition to Feshbach molecules:} Left and right column
represent spins of two atoms with different momentum. (a) and (b)
represent two cases in the absence (a) or in the presence (b) of SO
coupling. Red arrows represent spin direction at $t=0$ and the yellow
arrows represent spin direction at finite time $t$. \label{Fig1} }
\end{figure}

Let us consider two atoms on the positive scattering length
($a_s>0$) side of an $s$-wave Feshbach resonance.
In such a system, the Feshbach molecule
is in the singlet state $|S\rangle\equiv (\left|\uparrow\right\rangle_1\left|\downarrow\right\rangle_2-\left|
\downarrow\right\rangle_1\left|\uparrow\right\rangle_2)/\sqrt{2}$.
We assume
two fermionic atoms are initially prepared in the same spin state (say $\left|\downarrow\right\rangle$), with different momenta ${\bf p}$ and ${\bf
q}$, as represented by blue arrows in Fig. \ref{Fig1}. The initial
state under anti-symmetrization is given by
\begin{equation}
\left|\Psi\right\rangle_{i}=\frac{1}{\sqrt{2}}\left(\left|{\bf p}\right\rangle_1\left|{\bf
q}\right\rangle_2-\left|{\bf q}\right\rangle_1\left|{\bf
p}\right\rangle_2\right)\left|\downarrow\right\rangle_1\left|\downarrow\right\rangle_2.
\end{equation}
Now let us turn on a single particle term that couples two spin
states as
\begin{equation}
{\bf h}({\bf k})\cdot\boldsymbol{\sigma}. \label{couple}
\end{equation}
If the ``effective magnetic field" ${\bf h}$ is ${\bf k}$-independent,
it represents the case without SO coupling. In that case, ${\bf h}$ acts as
a uniform magnetic field and the two atoms
with different momentum always rotate in the same way. Therefore, at
a given time $t$, both of them rotate to the same direction
$|\hat{n}\rangle$, as shown in Fig. \ref{Fig1}(a). The final state
wave function is then given by
\begin{equation}
|\Psi\rangle_{f}=\frac{1}{\sqrt{2}}\left(|{\bf p}\rangle_1|{\bf
q}\rangle_2-|{\bf q}\rangle_1|{\bf
p}\rangle_2\right)|\hat{n}\rangle_1|\hat{n}\rangle_2.
\end{equation}
Since this state remains in triplet channel,
$\langle S|\Psi\rangle_f$ is always zero. Thus, there cannot be any
coherent transition to the Feshbach molecular state. On the other hand, if any component of
${\bf h}$ depends on ${\bf k}$, it means that the spin and momentum
are coupled. In this case, the amount of rotation each spin executes depends on
its momentum and is in general  different for different momentum. 
Suppose at time $t$, atom with momentum ${\bf
p}$ rotates to $|\hat{n}_{\bf p}\rangle$ and atom with momentum
${\bf q}$ rotates to $|\hat{n}_{\bf q}\rangle$, as shown in Fig.
\ref{Fig1}b, the final state wave function can now be written as
\begin{equation}
|\Psi\rangle_{f}=\frac{1}{\sqrt{2}}\left(|{\bf p}\rangle_1|{\bf
q}\rangle_2|\hat{n}_{\bf p}\rangle_1|\hat{n}_{\bf q}\rangle_2-|{\bf
q}\rangle_1|{\bf p}\rangle_2|\hat{n}_{\bf q}\rangle_1|\hat{n}_{\bf
p}\rangle_2\right) \label{wf}.
\end{equation}
It is straightforward to show the wave function Eq. (\ref{wf}) can
be rewritten as
\begin{eqnarray}
|\Psi\rangle_{f}=\frac{\left(|{\bf p}\rangle_1|{\bf
q}\rangle_2-|{\bf q}\rangle_1|{\bf
p}\rangle_2\right)}{2}|\widetilde{T}\rangle+\frac{\left(|{\bf
p}\rangle_1|{\bf q}\rangle_2+|{\bf q}\rangle_1|{\bf
p}\rangle_2\right)}{2}|\widetilde{S}\rangle \nonumber
\end{eqnarray}
where $|\widetilde{T}\rangle= (|\hat{n}_{\bf p}\rangle_1|\hat{n}_{\bf
q}\rangle_2+|\hat{n}_{\bf q}\rangle_1|\hat{n}_{\bf
p}\rangle_2)/\sqrt{2}$ and $|\widetilde{S}\rangle= (|\hat{n}_{\bf
p}\rangle_1|\hat{n}_{\bf q}\rangle_2-|\hat{n}_{\bf
q}\rangle_1|\hat{n}_{\bf p}\rangle_2)/\sqrt{2}\propto|S\rangle$ are
triplet and singlet components, respectively. Thus, $\langle
S|\Psi\rangle_f$ is non-zero and these two atoms can experience
$s$-wave resonant interaction. Therefore, a transition to Feshbach molecular
state can be induced.

Nevertheless, in reality, it is known that a momentum-independent coupling, such as radio-frequency (rf) coupling, can also produce Feshbach molecules in a degenerate Fermi gas. In such a process the atoms in
$|\downarrow\rangle$ first evolve to the superposition of
$|\uparrow\rangle$ and $|\downarrow\rangle$ through rf coupling.
Then, after decoherence, it becomes an incoherent
mixture of scattering atoms in $|\uparrow\rangle$ and $|\downarrow\rangle$, and further, the inelastic collision can bring
some atoms in the mixture into molecules~\cite{molecules}.
It is important to notice that decoherence process has to be involved in such a process. While in contrast, the SO-coupling-induced transition discussed above does not require any incoherent process, and is a fully \textit{quantum coherent} process.

In our experiment, a pair of $772.4$ nm Raman lasers is applied to
spin polarized ${}^{40}$K gas in $|F,m_{\rm F}\rangle= |9/2,-9/2\rangle$ state.
Two Raman beams are linearly polarized along $\hat{z}$ and $\hat{y}$
axis, respectively, which correspond to $\pi$ and $\sigma$ polarization along the
quantization axis $\hat{z}$. Thus, the Raman process couples
$|9/2,-9/2\rangle$ (denoted by $\left|\downarrow\right\rangle$) to
$|9/2,-7/2\rangle$ (denoted by $\left|\uparrow\right\rangle$), and
the momentum transfer in the Raman process is $2k_0=2k_{\text
r}\sin(\theta/2)$, where $k_{\text r}=2\pi \hbar/\lambda$ is the
single-photon recoil momentum, $\lambda$ is the wavelength of the
Raman beam, and $\theta$ is the angle between two Raman beams.
The recoil energy $E_{\text{r}}= k_{\text{r}}^{2}/2m = h\times 8.36$ kHz. The relative
frequency between two lasers $\omega_1-\omega_2$ can be precisely
controlled, and the detuning $\delta\equiv\hbar(\omega_1-\omega_2)-\hbar\omega_{\rm Z}$,
where $\hbar\omega_{\rm Z}$ is the Zeeman splitting between
$\left|\uparrow\right\rangle$ and $\left|\downarrow\right\rangle$. This Raman process is
described by single particle Hamiltonian~\cite{spielman}
\begin{equation}
\hat{H}_0=\frac{(p_x-k_0\sigma_z)^2}{2m}+\frac{\Omega}{2}\sigma_x-\frac{\delta}{2}\sigma_z+\frac{k^2_y+k^2_z}{2m}.
\end{equation}
Here, $p_{x}$ denotes the quasi-momentum of atoms, which relates to
the real momentum $k_{x}$ as $k_{x}= p_{x}\mp k_{0}$ with $\mp$ for
spin-up and down, respectively, and $\Omega$ is the strength of the
Raman coupling. Comparing with Eq. (\ref{couple}), it is clear that
${\bf h}=(\Omega/2, 0, -\delta/2-p_x k_0/m)$. If two Raman beams are
parallel to each other, we have $\theta=0$ and thus $k_0=0$. In this
case there is no SO coupling. When $\theta\neq 0$, $k_0$ becomes
non-zero and there will be SO coupling effect. According to above
analysis, a fully polarized Fermi gas cannot be coupled to Feshbach
molecular state if two Raman beams are parallel, while coherent
molecule production is allowed if they are not parallel.

\begin{figure}[tbp]
\centerline{
\includegraphics[bb=75 400 353 608,clip,width=7cm]{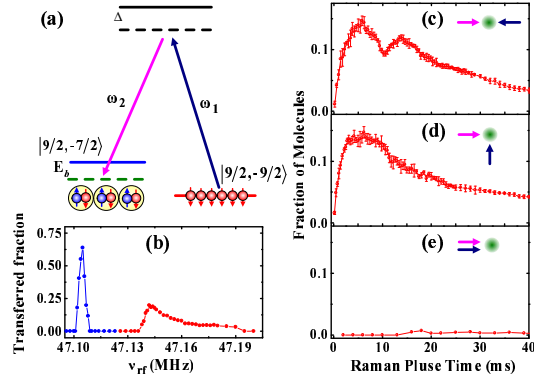}
} 
\caption{(Color online). \textbf{Energy level diagram and spin-orbit
coupling induced Feshbach molecules}. \textbf{(a)} Schematic diagram
of the energy levels. A pair of Raman lasers couples spin polarized
state $|9/2,-9/2\rangle$ to Feshbach molecules in Fermi gases
${}^{40}$K. \textbf{(b)} Radio-frequency spectrum
$|9/2,-7/2\rangle$ to $|9/2,-5/2\rangle$ transition applied to a
mixture of Feshbach molecules and scattering atoms in $|9/2,-7/2\rangle$.
\textbf{(c-e)} The population of Feshbach molecules detected by the
rf pulse as a function of duration time of the Raman pulse. The
angle of two Raman beams is $\theta=180^{\circ}$ (c), $\theta=90^{\circ}$
(d) and $\theta=0^{\circ}$ (e). The Raman coupling strength is
$\Omega=1.3E_{r}$ and the two-photon Raman detuning is
$\delta=-E_b=-3.59E_{\text{r}}$. }
\label{Fig2}
\end{figure}

Our experiment is performed at $201.4$ G,  below the Feshbach
resonance between $|9/2,-9/2\rangle$ and $|9/2,-7/2\rangle$ located
at $202.2$ G, which corresponds to a binding energy of $E_b=h\times30$
kHz (corresponding to 3.59$E_{\text{r}}$) for the Feshbach molecules and
$1/(k_\text{F}a_\text{s})\approx 0.92$ for our typical density.
After applying the Raman lasers for certain duration time, we turn
off the Raman lasers and measure the population of Feshbach molecule and
atoms in $|9/2,-7/2\rangle$ state with a radio-frequency (rf) pulse.
This rf field drives a transition from $|9/2,-7/2\rangle$ to
$|9/2,-5/2\rangle$. After the rf pulse, we abruptly turn off the
optical trap and the magnetic field, and let atoms ballistically
expand for $12$ ms in the presence of a magnetic field gradient applied
along $\hat{y}$, and finally take absorption image along $\hat{z}$
to measure the population of $|9/2,-5/2\rangle$ state. For a mixture
of $|9/2,-7/2\rangle$ and Feshbach molecules, as a function of rf
frequency $\nu_\text{rf}$, we find two peaks in the population of
$|9/2,-5/2\rangle$, as shown in Fig. \ref{Fig2}(b). The first peak
(blue curve) is attributed to free atom-atom transition and the
second peak (red curve) is attributed to molecule-atom transition.
Thus, in the following, we set $\nu_\text{rf}$ to $47.14$ MHz to
measure Feshbach molecules.

When the two-photon Raman detuning $\delta$ is set to
$\delta=-E_b=-3.59 E_{\text{r}}$~\cite{bound}, as shown in
Fig. \ref{Fig2}(a), we measure the population of Feshbach molecule
as a function of duration time for three different angles,
$\theta=180^{\circ}$, $\theta=90^{\circ}$, and $\theta=0^{\circ}$, as shown in
Fig. \ref{Fig2}(c), (d) and (e). We find for $\theta=180^{\circ}$,
Feshbach molecules are created by Raman process and the coherent
Rabi oscillation between atom-molecule can be seen clearly. For
$\theta=90^{\circ}$, production of Feshbach molecules is reduced a
little bit and the atom-molecule Rabi oscillation becomes invisible. For
$\theta=0^{\circ}$, no Feshbach molecule is created even up to $40$ ms,
which means the transition between Feshbach molecules and a
fully polarized state is prohibited if Raman process has no momentum
transfer.

\begin{figure}
\centerline{
\includegraphics[bb=42 403 425 740,clip,width=5.5cm]{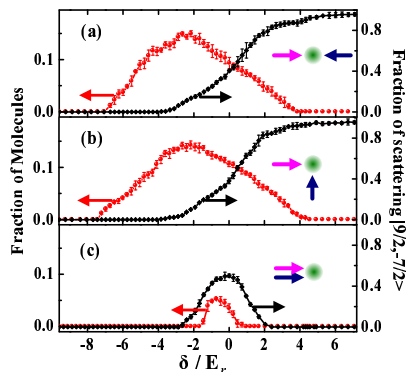}
} 
\caption{(Color online). \textbf{The population of Feshbach
molecules and scattering atoms in $|9/2,-7/2\rangle$ state as a
function of two-photon detuning of the Raman pulse}. The Raman
coupling strength is $\Omega=1.3E_{\text{r}}$ and the duration of
the Raman pulse is $15$ ms.  Angle of two Raman beams is
$\theta=180^{\circ}$(a), $\theta=90^{\circ}$(b), and
$\theta=0^{\circ}$(c), respectively. The red data points are
Feshbach molecular population and the black data points are
population of scattering atoms in  $|9/2,-7/2\rangle$ state.
\label{Fig3} }
\end{figure}

Fig. \ref{Fig3} shows the population of Feshbach molecules detected
by the rf pulse as a function of two-photon detuning $\delta$ with the
fixed Raman coupling strength $\Omega=1.3E_{r}$ and
the pulse duration $15$ ms. For $\theta=180^{\circ}$
and $\theta=90^{\circ}$ we find that the formation of Feshbach molecules starts
to appear when $\delta\gtrsim -7.18 E_{\text{r}}$, reaches a
maximum around $\delta\approx -2.39 E_{\text{r}}$
 (it is a little bit larger than $-E_b=-3.59E_{\text{r}}$, which
is probably due to the momentum recoil from the Raman beams
), and gradually decreases to zero around
$\delta=+3.59E_{\text{r}}$, as shown by red data points in Fig.
\ref{Fig3}(a) and (b). While for $\theta=0^{\circ}$, we find no
Feshbach molecule production until $\delta\gtrsim -1.79E_{\text{r}}$
and a Feshbach molecular population with narrower width is only near
$\delta\sim 0$, as shown in Fig. \ref{Fig3}(c). The peak value in
(c) is also much reduced compared to (a) and (b).

The atom-molecule transition shown in Fig.~\ref{Fig3} contains
both the SO-coupling-induced coherent process and the incoherent process
discussed above. For the incoherent process, sufficient population of
scattering atoms in $|9/2,-7/2\rangle$  is
inevitable. In contrast, the coherent process can still exist even when the population of the scattering atoms in $|9/2,-7/2\rangle$ is negligible at sufficient large detunning.
Thus, to further distinguish these two processes, we measure the
population of scattering atoms in $|9/2,-7/2\rangle$ state for three cases with
$\theta=180^{\circ}\ {\rm (a)},\ 90^{\circ} {\rm (b)},\ 0^{\circ}\ {\rm (c)}$, after a Raman pulse with the same intensity
and the same duration. We find in all the three cases, the population
of scattering atoms in $|9/2,-7/2\rangle$ becomes non-zero for
$\delta\gtrsim-3.59 E_{\text{r}}$ as shown in black data points in
Fig. \ref{Fig3}. The difference between (a,b) and (c) is that
for (a,b), the system mainly populate in $|9/2,-7/2\rangle$ when
$\delta>3.59 E_{\text{r}}$, while for (c) the population of
$|9/2,-7/2\rangle$ vanishes when $\delta>2.39 E_{\text{r}}$. This
is because for the case without SO coupling (c), the resonance
always takes place when two-photon detuning matches Zeeman energy,
i.e. $\delta=0$, for atoms in all momentum. While for (a) and (b)
with SO coupling, the resonance frequency is momentum dependent and
spend over a much wider frequency range~\cite{Jing,MIT,Jing_Raman1,Jing_Raman2}.

Comparing the populations of scattering atoms in the state $|9/2,-7/2\rangle$ and
Feshbach molecules
in Fig. \ref{Fig3}, we find that for $\theta=180^{\circ}$ and
$\theta=90^{\circ}$ (SO coupling case), significant molecule population
exists in the frequency regime $\delta\lesssim -3.59 E_{\text{r}}$
where almost no scattering atoms in $|9/2,-7/2\rangle$ are found. This
confirms the coherent nature of Feshbach molecular production. On the other hand,
for $\theta=0^{\circ}$ (no SO coupling), no Feshbach molecule can be found where no population
of scattering atoms in $|9/2,-7/2\rangle$ atoms can be seen (either
$\delta\lesssim -2.39 E_{\text{r}}$ or $\delta\gtrsim +2.39 E_{\text{r}}$). This shows
that decoherence process is key in producing scattering atoms in the state
$|9/2,-7/2\rangle$, which is prerequisite for incoherent molecular formation~\cite{molecules}.


\begin{figure}
\centerline{
\includegraphics[bb=25 413 555 760,clip,width=7cm]{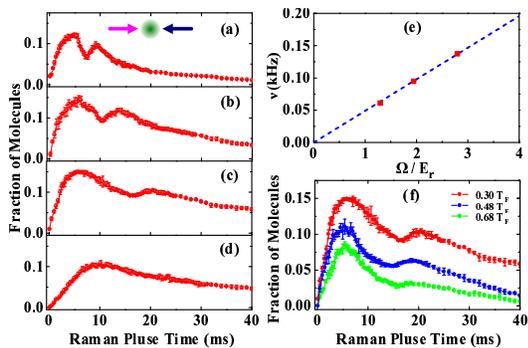}
} 
\caption{(Color online). \textbf{The dependence on the Raman
coupling strength and temperature for coherent Rabi oscillations
between a spin polarized Fermi gas and Feshbach molecular state}.
(a-e). Population of Feshbach molecules as a function of duration
time of Raman laser. The angle of two Raman beams is
$\theta=180^{\circ}$ and the two-photon detuning is set as
$\delta=-E_b=-3.59E_{\text{r}}$. For (a-d), temperature $T/T_\text{F}=0.3$, and
the Raman coupling strength is $\Omega=2.8E_{\text{r}}$ (a),
$\Omega=1.95E_{\text{r}}$ (b), $\Omega=1.3E_{\text{r}}$ (c), and
$\Omega=0.65E_{\text{r}}$ (d), respectively. (e) For $T/T_\text{F}=0.3$,
the Rabi frequencies obtained from (a-c) as the function of the
Raman coupling strength. (f), Feshbach molecular fractions as a function of pulse time
for different temperatures. The Raman
coupling strength $\Omega=1.3E_{r}$. $T/T_{\text{F}}=0.3$ for
red curve, $T/T_{\text{F}}=0.48$ for blue curve, and
$T/T_{\text{F}}=0.68$ for green curve. \label{Fig4} }
\end{figure}

A more direct evidence for the coherent nature of molecular production
is the Rabi oscillation between Feshbach molecular state and a fully polarized Fermi gas.
Previously coherent atom-molecule oscillation has only been observed in bosonic
atomic gas~\cite{wieman} and boson-fermion mixture~\cite{Jin}. In a
Fermi gas the energy of atoms in scattering states spread over a
wide energy range of the order of Fermi energy (2.2$E_{\text{r}}$ in
our system), which inevitably leads to damping of Rabi oscillation. However, by
tuning the laser intensity and as a result, the magnitude of the Rabi frequency,
the oscillation period can be made shorter compared with damping time and can
be readily observed in the experiment. In Fig. \ref{Fig4} we
plot Feshbach molecular fraction as a function of duration time of Raman laser
with two-photon detuning tuned to molecule binding energy. In Fig.
\ref{Fig4}(a-c) at least one oscillation period can be identified.
While in case (d) with a smaller Raman intensity, oscillation
becomes invisible. For (a-c) we take the first minimum as one period
$\tau$, and plot Rabi oscillation frequency $\nu=1/\tau$ as a function of
Raman coupling $\Omega$ in Fig. \ref{Fig4}(e), and find a perfect
linear relation. This is indicative of a coherent process
in which the oscillation frequency is proportional to Raman-coupling
strength. Finally in Fig. \ref{Fig4}(f) we plot Feshbach
molecular fraction for various temperatures. We find when temperature
increases from $T/T_\text{F}=0.3$ to $T/T_\text{F}=0.68$, the
oscillation period is almost unchanged but the oscillation itself
becomes less and less visible. This shows the increase of damping
rate with the increase of temperature.

\begin{figure}
\centerline{
\includegraphics[bb=38 508 409 790,clip,width=7cm]{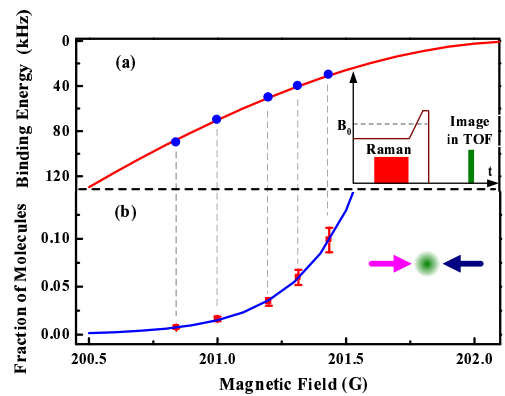}
} \vspace{0.1in}
\caption{(Color online). \textbf{The population of Feshbach
molecules as a function of the magnetic field}. The Raman coupling
strength is $\Omega=1.3E_{\text{r}}$ and the duration of the pulse
is $7$ ms. Angle of two Raman beams is $\theta=180^{\circ}$. The
detuning is chosen such that $\delta=-E_{\rm b}$ for each magnet
fields. (a) The Feshbach molecular binding energy as a function of
magnetic field. Inset of (a) shows the experimental sequence. (b)
Fraction of Feshbach molecules produced after the Raman pulse. The
Feshbach molecules are dissociated by a magnetic ramp over the
Feshbach resonance.
 \label{Fig5} }
\end{figure}

Finally, we study the creation of Feshbach molecules for different magnetic fields
corresponding to different binding energies $E_{b}$ of the molecules.
The detuning $\delta$ is chosen to be $\delta=-E_{b}$. Here, the
Feshbach molecules are dissociated by a magnetic sweep over the
Feshbach resonance instead of the rf pulse. As shown in Fig.~\ref{Fig5},
we find that the Feshbach molecular population
increases at the higher magnetic field, or lower binding energy $E_b$.
This is because the atom-molecule transition amplitude
depends on the overlap between the wave-function of Feshbach molecule and the one of two free atoms
(i.e., the Franck-Condon factor), which increases with $E_b$.
%
%

In conclusion, by applying SO coupling, we have demonstrated
coherent production of Feshbach molecules from a fully polarized
Fermi gas, as well as a coherent oscillation between them, provided
 that the coupling strength is strong enough. Such a coherent process
 reveals that, in the presence of SO coupling, the atomic triplet state is coupled
 to the singlet state, and thus
 the bound state of a system with SO coupling contains both singlet and triplet components. For symmetry
reason, the triplet component should at least be a $p$-wave pairing.  Although
the temperature of our current system is still above the
condensation temperature of these pairs, one may still expect some
interesting physics of these noncondensed nontrivial pairs. At low
temperature and lower dimension these pairs will exhibit topological
superfluidity.

\textbf {Method:} This experiment starts with a degenerate Fermi gas
of about $2\times10^{6}$ ${}^{40}$K in the $|9/2,9/2\rangle$
state, which has been evaporatively cooled to $T/T_{F}\approx0.3$
with bosonic ${}^{87}$Rb atoms in the $|2,2\rangle$ inside
the crossed optical trap ~~\cite{Jing,four2,four3,four4}, where
$T_{F}$ is the Fermi temperature defined by $T_{F}=(6
N)^{1/3}\hbar\overline{\omega}/k_{B}$, and
$\overline{\omega}\simeq2\pi\times80$ Hz in our system, $N$ is the
number of fermions. A $780$ nm laser pulse is applied for $0.03$ ms
to remove the ${}^{87}$Rb atoms in the mixture without heating
${}^{40}$K atoms. Subsequently, the fermionic atoms are transferred
into the lowest state $|9/2,-9/2\rangle$ via a rapid
adiabatic passage induced by a radio-frequency field of $80$ ms at
$4$ G. A homogeneous bias magnetic field for magnetic Feshbach
resonance along the $z$ axis (gravity direction) is produced by the
quadrupole coils (operating in the Helmholtz configuration).

A pair of $772.4$ nm Raman laser are extracted from a CW Ti-sapphire
single frequency laser. Two Raman beams are frequency-shifted around
$-77$ MHz and $-122$ MHz by two single-pass acousto-optic modulators
(AOM), respectively, to precisely control their frequency
difference. These two Raman beams has a maximum intensity $I= 130$
mW for each beam, and they overlap in the atomic cloud with
$1/e^{2}$ radii of 200 $\mu m$.

\begin{acknowledgments}
We would like to thank Cheng Chin for helpful discussions. This
research is supported by National Basic Research Program of China
(Grant No. 2011CB921601, No. 2011CB921500), NSFC (Grant No.
11234008, No.11004118 and No. 11174176), NSFC Project for Excellent
Research Team (Grant No. 61121064). PZ would also like to thank the
NCET Program for support. SZ is supported by the start-up fund from
the University of Hong Kong.
\end{acknowledgments}


\begin{thebibliography}{99}
\bibitem{bound} At the same time we should be careful that each single beam $\omega_1$ and $\omega_2$ will not cause bound-to-bound transition.
\end{thebibliography}
\end{document}